\documentclass[%
 reprint,
superscriptaddress,
 amsmath,amssymb,
 aps,
]{revtex4-2}

\usepackage{graphicx}
\usepackage{dcolumn}
\usepackage{bm}
\usepackage{hyperref}
\hypersetup{
    pdfnewwindow=true,      
    colorlinks=true,        
    linkcolor=blue,         
    citecolor=blue,        
    filecolor=blue,         
    urlcolor=blue           
}

\newcommand*{\add}[1]{\textcolor{black}{#1}}
\newcommand*{\addb}[1]{\textcolor{black}{#1}}

\newcommand{\aap}{Astron. Astrophys.}

\newcommand{\apjl}{Astrophys. J. Lett.}
\newcommand{\cqg}{Class. Quantum Grav.}

\newcommand{\mnras}{Mon. Notices Royal Astron. Soc}

\begin{document}


\title{Hierarchical Black Hole Mergers in Active Galactic Nuclei}

\author{Y. Yang}
\affiliation{Department of Physics, University of Florida, PO Box 118440, Gainesville, FL 32611-8440, USA}
\author{I. Bartos}
\thanks{imrebartos@ufl.edu}
\affiliation{Department of Physics, University of Florida, PO Box 118440, Gainesville, FL 32611-8440, USA}
\author{V. Gayathri}
\affiliation{Indian Institute of Technology Bombay, Powai, Mumbai 400 076, India}
\affiliation{Department of Physics, University of Florida, PO Box 118440, Gainesville, FL 32611-8440, USA}
\author{K.E.S. Ford}
\affiliation{Department of Science, CUNY-BMCC, 199 Chambers St., New York NY 10007}
\affiliation{Department of Astrophysics, American Museum of Natural History, Central Park West, New York, NY 10028}
\affiliation{Physics Program, The Graduate Center, CUNY, New York, NY 10016}
\author{Z. Haiman}
\affiliation{Department of Astronomy, Columbia University in the City of New York, 550 W 120th St., New York, NY 10027, USA}
\author{S. Klimenko$^*$}
\affiliation{Department of Physics, University of Florida, PO Box 118440, Gainesville, FL 32611-8440, USA}
\author{B. Kocsis}
\affiliation{E\"otv\"os University, Institute of Physics, P\'azm\'any P. s. 1/A, Budapest, 1117, Hungary}
\author{S. M\'arka}
\affiliation{Department of Physics, Columbia University in the City of New York, 550 W 120th St., New York, NY 10027, USA}
\author{Z. M\'arka}
\affiliation{Columbia Astrophysics Laboratory, Columbia University in the
 City of New York, 550 W 120th St., New York, NY 10027, USA}
\author{B. McKernan}
\affiliation{Department of Science, CUNY-BMCC, 199 Chambers St., New York NY 10007}
\affiliation{Department of Astrophysics, American Museum of Natural History, Central Park West, New York, NY 10028}
\affiliation{Physics Program, The Graduate Center, CUNY, New York, NY 10016}
\author{R. O'Shaughnessy}
\affiliation{Rochester Institute of Technology, Rochester, NY 14623, USA}

\begin{abstract}
The origins of the stellar-mass black hole mergers discovered by LIGO/Virgo are still unknown. Here we show that, if migration traps develop in the \add{accretion} disks of Active Galactic Nuclei (AGNs) and promote the mergers of their captive black holes, the majority of black holes within disks will undergo hierarchical mergers---with one of the black holes being the remnant of a previous merger. 40\% of AGN-assisted mergers detected by LIGO/Virgo will include a black hole with mass $\gtrsim 50$\,M$_\odot$, the mass limit from stellar core collapse. Hierarchical mergers at traps in AGNs will exhibit black hole spins (anti-)aligned with the binary's orbital axis, \add{a distinct property from other hierarchical channels}. \addb{Our results are suggestive, although not definitive (with Odds ratio of $\sim 1$), that LIGO's heaviest merger so far, GW170729, could have originated from this channel.}
\end{abstract}

\maketitle



{\it Introduction.---} The number of binary black hole (BBH) mergers detected by Advanced LIGO \cite{2015CQGra..32g4001L} and Advanced Virgo \cite{2015CQGra..32b4001A} is rapidly growing. More than ten mergers have been discovered during LIGO/Virgo's first two observing runs \cite{2018arXiv181112907T,2019arXiv190407214V}, and many more are expected in the current third observing run and beyond \cite{2018LRR....21....3A}.

Despite the growing number of observations, the formation mechanism of the detected BBHs is currently not understood. Favored scenarios include isolated binary evolution in which the black holes (BHs) are produced in a binary star system \cite{2010ApJ...714.1217B,Postnov2014, 2016MNRAS.460.3545D}, and dynamical formation in which the BHs become gravitationally bound following a chance encounter in a dense stellar environment such as galactic nuclei or globular clusters \cite{2006ApJ...645L.133H,2009MNRAS.395.2127O,Benacquista2013,2013MNRAS.432.2779B,2016PhRvD..93h4029R,2016ApJ...824L...8R,2018PhRvL.120o1101R,2016ApJ...831..187A,2016MNRAS.458.1450W}.

AGNs represent a unique environment that can assist and alter the evolution of BBH mergers. The nuclei of active galaxies is expected to harbor potentially tens of thousands of stellar-mass BHs that moved into the innermost parsec due to mass segregation \cite{1993ApJ...408..496M,2000ApJ...545..847M,2014ApJ...794..106A,2018Natur.556...70H,2018MNRAS.478.4030G}. Interaction with the AGN \add{accretion} disk \add{(hereafter AGN disk)} can align the orbits of these BHs with the disk \cite{1991MNRAS.250..505S,1993ApJ...409..592A,2004ApJ...608..108G,2012MNRAS.425..460M,2014MNRAS.441..900M,2017ApJ...835..165B}. Alternatively, some BHs can be formed in the disk itself \cite{2007MNRAS.374..515L,2017MNRAS.464..946S}. Once in the disks, interaction between the rotating gas may move the BHs to migration traps within the disk, to about 300 Schwarzschild radii from the central supermassive BH (SMBH; \cite{2012MNRAS.425..460M,2016ApJ...819L..17B,2018arXiv180702859S}). If multiple BHs move into the disk, they will eventually meet in the migration trap and merge. This merger will be rapid due to dynamical friction within the disk \cite{2017ApJ...835..165B,1993Natur.364..421K,1993Natur.364..423S,2002ApJ...576..899P,2011ApJ...726...28B}. Alternatively, BBHs can also align their orbit with the disk, and merge rapidly in the disk without reaching the migration trap \cite{2017ApJ...835..165B}.

AGN-assisted BH mergers have distinct properties that could differentiate them from other formation channels. These include their mass distribution \add{in which heavier BHs are expected to be overrepresented by a factor roughly proportional to their mass} \cite{2019ApJ...876..122Y}, their location in AGNs that can be differentiated from binaries formed in other types of galaxies \cite{2017NatCo...8..831B,2019arXiv190202797C}, possible electromagnetic signatures produced due to the BHs accreting from the surrounding dense gas \cite{2017ApJ...835..165B,2017MNRAS.464..946S}, or center-of-mass acceleration \cite{2015MNRAS.452L...1M,2017ApJ...834..200M,2017PhRvD..96f3014I,2018MNRAS.475.4595W}.

In addition, as multiple BHs align their orbits with the AGN disk and move to the migration trap, they merge and remain near the migration trap, enabling the remnant to merge with additional BHs \cite{2012MNRAS.425..460M}. Such hierarchical mergers will lead to distinct, high BH masses and characteristic spin properties that can be identified via gravitational wave observations \cite{2017PhRvD..95l4046G,2019arXiv190605295G}.

Here we examined the prevalence and observational signatures of hierarchical mergers in AGN disks. We carried out Monte Carlo simulations of BH orbital alignments and mergers in a population of AGNs, taking into account the possibility of hierarchical mergers during the lifetime of AGNs. We computed the resulting BH mass and spin distributions. Finally, we compare these distributions with the heaviest BH merger detected by LIGO and Virgo, GW170729 \cite{2018arXiv181112907T}.



{\it Stellar-mass black hole population.---} Our simulations follow the method of \cite{2019ApJ...876..122Y}, who semi-analytically calculated the interaction between a stellar-mass BH orbiting a SMBH and the AGN disk. They carried out a Monte Carlo simulation with a parametrized power-law cusp distribution of stellar-mass BHs around SMBHs, and a realistic distribution of SMBH masses and AGN disk properties. Here we consider their result for the fiducial BH mass distribution $dN/dm_{\rm bh}\propto m_{\rm bh}^{-1}$ within the AGN disk, which they obtained for an initial mass function $dN/dm_{\rm bh}\propto m_{\rm bh}^{-2.35}$. The BHs had a thermal eccentricity distribution and isotropic directional distribution prior to alignment with the AGN disk. 

Initial BH masses were limited to $[5\mbox{M}_\odot,50\mbox{M}_\odot]$. The upper mass limit of $\sim 50\mbox{M}_\odot$ is due to pair-instability mass loss in stars that would otherwise produce heavier BHs \cite{2017ApJ...836..244W,2016A&A...594A..97B,2018MNRAS.480.2011G}.


{\it Migration and merger time frames.---} Migration traps have been proposed to develop in AGN disks in analogy with those previously invoked for protoplanetary disks (e.g. \cite{2012MNRAS.425..460M,2016ApJ...819L..17B,2018arXiv180702859S}). While their existence is not yet certain, if they develop they will attract BHs from within the disk over a characteristic time frame of $10^5$\,yr \cite{2012MNRAS.425..460M}. 

Following the calculations of \cite{2017ApJ...835..165B}, we find that the merger time for a 30\,M$_\odot-30\,$M$_\odot$ binary in a migration trap of a 10$^{6}$\,M$_\odot$ SMBH accreting at $\dot{m}_{\bullet}=0.1$ will be about $10^5$\,yr. \add{Therefore, as the time of migration and merger for BHs are much shorter than typical orbital alignment times with the AGN disk, we neglect them in our Monte Carlo simulation discussed below.}

We note that mergers may occur prior to arrival at a trap \cite{2019arXiv190704356M}, which would lead to less higher-generation mergers. It is also possible that migration traps do not develop. Migration traps have been proposed to occur analogously to protoplanetary disks, coinciding with local surface density turning points, where torques on migrating compact objects vanish \cite{2012MNRAS.425..460M,2016ApJ...819L..17B}. These initial estimates use a simplified disk model, e.g. excluding the impact of migrators on the local disk density. The lack of a trap would lead to less or no higher-generation mergers in AGN disks. Therefore, hierarchical BH mergers are also a test of the existence of migration traps and the merger process.


{\it Simulation of hierarchical mergers.---} For each AGN disk we generated a population of mergers. Hereafter, the merger of two BHs that each came from the initial population (presumably from stellar evolution) will be referred to as first-generation, or 1g. The merger of a BH that is the remnant of a 1g merger with another BH will be referred to as a second-generation, or 2g, merger. We define 3g, 4g, etc. similarly. We only consider mergers in which at least one of the BHs is {\it not} the result of a previous merger. This is expected for AGN disks if single BHs move into the disk, since the migration and merger rate is much faster than the characteristic time difference between two BHs moving into the disk. One exception is when BBHs migrate into the disk, which can lead to both BHs being at least 2g. For simplicity we ignore this possibility here.  

The masses and spins of the BHs formed in mergers were calculated using the {\it surfinBH} package for mass ratios $m_2/m_1>0.1$ \cite{2019PhRvL.122a1101V}. For $m_2/m_1\leq0.1$, we used the results of \cite{2012ApJ...758...63B} and \cite{2016ApJ...825L..19H} to calculate the final mass and spin, respectively. We characterized the BH's spins with the binary' effective spin 
\begin{equation}
\chi_{\rm eff}\equiv \frac{c}{GM}\left(\frac{\vec{S}_1}{m_1} + \frac{\vec{S}_2}{m_2}\right)\cdot\frac{\vec{L}}{|\vec{L}|}
\end{equation}
where $M=m_1+m_2$ is the total mass of the binary, $\vec{S}_{1,2}$ are the spin angular momentum vectors of the BHs in the binary, and $\vec{L}$ is the orbital angular momentum vector. This mass-weighted sum of the spins parallel to the binary orbit is the spin parameter that is the most accurately measured with gravitational waves. 

In the following we assume that all 1g BHs have zero spins. By investigating non-zero distributions, we found that our results do not depend significantly on this assumption, since the 1g spin direction is isotropically distributed and hence will have limited effect on $\chi_{\rm eff}$  (see \cite{2017Natur.548..426F}). \addb{Further, all but two of the BH mergers detected so far by LIGO/Virgo are consistent with zero BH spin (GW151226 and GW170729; \citealt{2018arXiv181112907T})}. We additionally assumed that accretion does not significantly alter the BHs' spin \addb{ (although see \cite{2019arXiv190908384Y}). For example an initially non-spinning BH that accretes at the Eddington rate (with efficiency $\epsilon=0.1$) for a full fiducial AGN lifetime of $10^7$\,yr will reach a dimensionless spin of 0.3 \cite{1970Natur.226...64B}.}

We assume all binary orbital axes to be aligned with the AGN disk, therefore the spins of higher-generation BHs will be either aligned or anti-aligned with newly formed binaries. 

In order to evaluate the fractions of each generation of mergers, we took into account that the number of BHs whose orbit is dragged into the AGN disk within the disks lifetime, taken to be $\tau_{\rm AGN}=10^7$\,yr \cite{2004cbhg.symp..169M}, has a Poisson distribution. We assumed that BHs on orbits aligned with the AGN disk migrate to the trap in the disk. Each new BH that reaches the migration trap merges with the BH already there. We assumed that merger remnants remain in the trap or quickly migrate back to it before the next BH reaches the trap. This is expected to be the case as natal kicks from the merger, which are on the order of several 100\,km\,s$^{-1}$, will not be able to substantially change the orbits, given typical orbital velocities of 20,000\,km\,s$^{-1}$. Small deviations quickly vanish due to orbital alignment.

As BHs in traps merge in sequence, the fraction of each generation are:
\begin{equation}
    P_{g}(\rm n)=\frac{1}{\lambda-1+e^{-\lambda}}\sum_{k=n+1}^{\infty}\mbox{Poiss}(\rm k,\lambda),\quad n=1,2,3...
\label{eq:generations}    
\end{equation}
where $\mbox{Poiss}$ is the Poisson distribution, and the term before the sum on the right side is a suitable normalization factor.

The expected value $\lambda$ is essentially independent of the mass of the SMBH, and weakly depends on the accretion rate (see Fig. 6 in \cite{2019ApJ...876..122Y}). Here we adopted a fiducial accretion rate of $\dot{m}=0.1$ onto the SMBH, with which we get $\lambda\sim2.5$.

\label{sec:results}

\begin{figure}
   \centering  
   \includegraphics[width=0.47\textwidth]{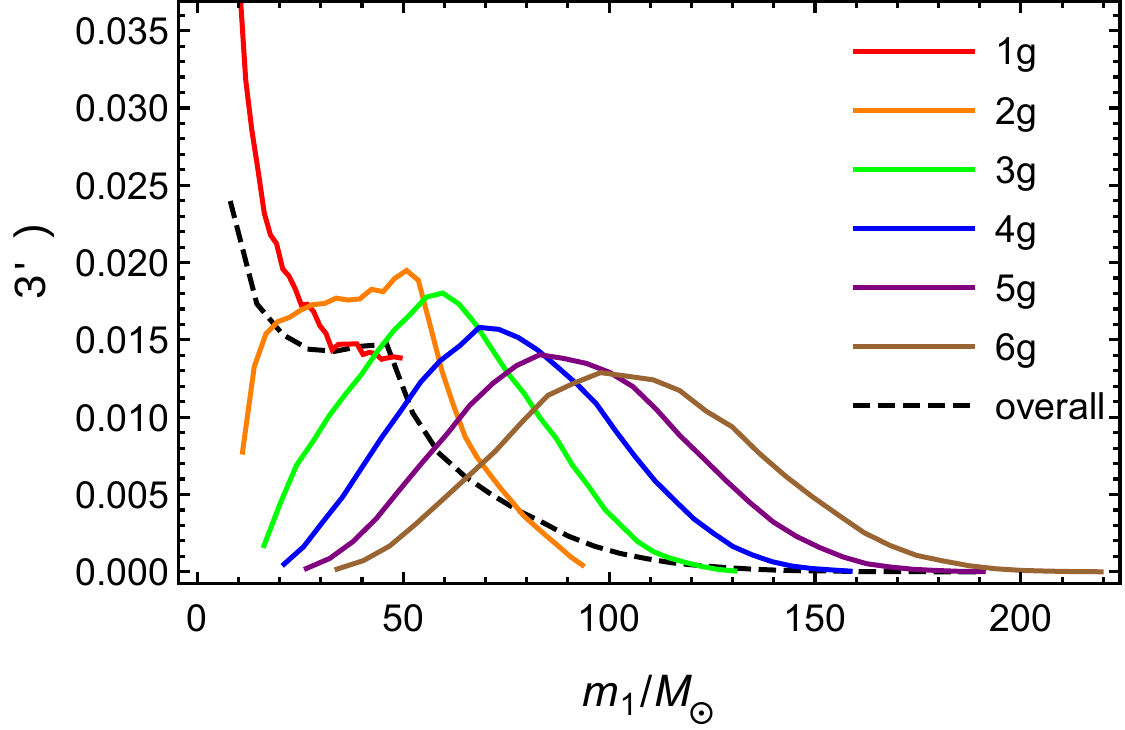}
   \caption{The distribution of $m_{1}$, we take into account the contribution from each generation of mergers. }   
\label{fig:mass}
\end{figure}

\begin{figure}
   \centering  
   \includegraphics[width=0.47\textwidth]{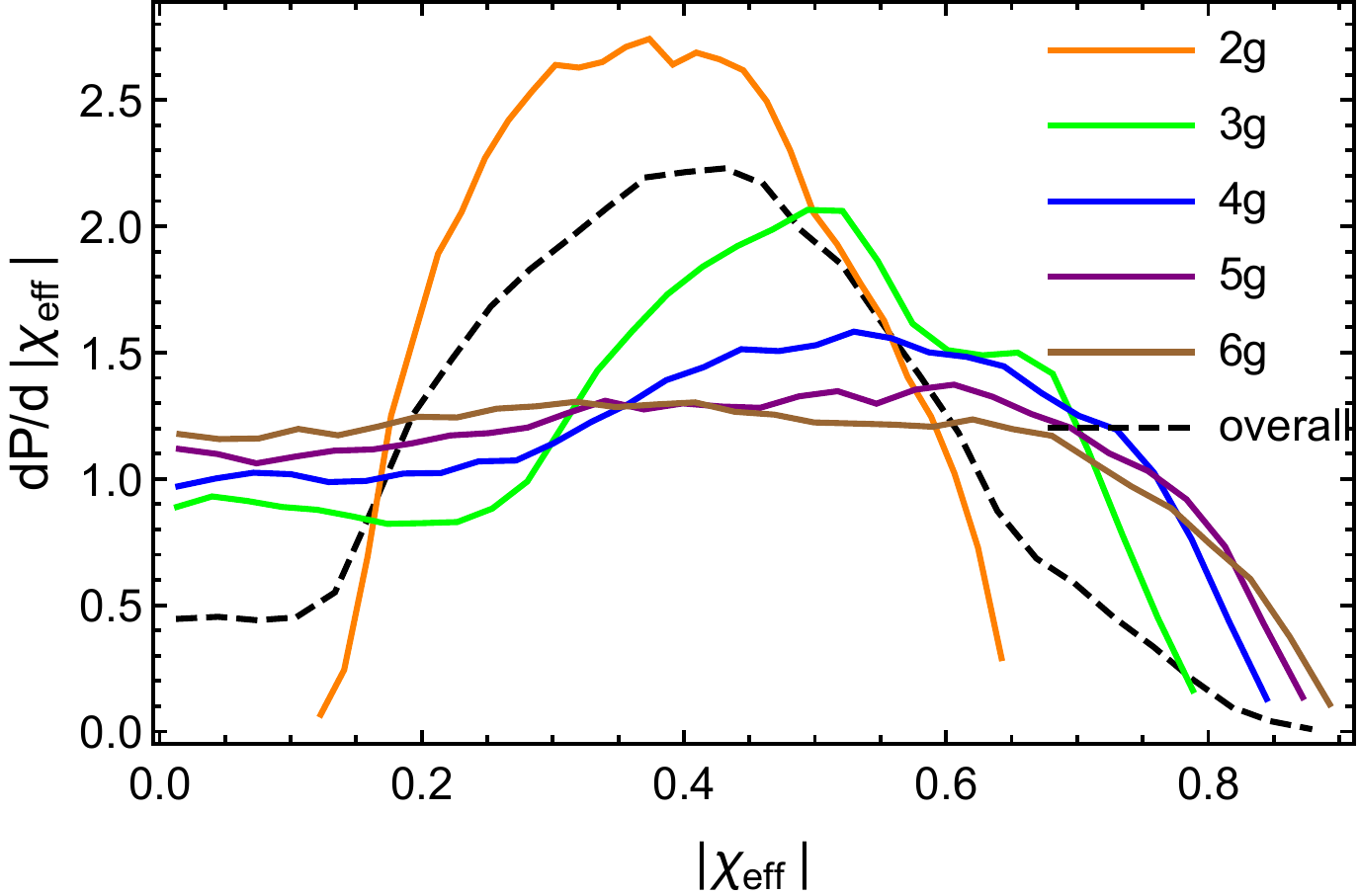}   
   \caption{The distribution of $\chi_{\rm eff}$ for different generations of mergers, and the overall distribution with all generations combined (see legend). For anti-aligned orbits $\chi_{\rm eff}$ is negative but is otherwise distributed identically to the shown distribution.}   
\label{fig:spin}
\end{figure}

{\it Fraction of hierarchical mergers.---} 
Using Eq. \ref{eq:generations} we found that the $\{47\%,29\%,15\%,6\%,1\%\}$ of AGN-assisted mergers are 1g, 2g,...5g, respectively. As each merger remnant is retained within the disk, higher-generation mergers are common. We find that the majority of mergers will be higher-generation. The prevalence of higher-generation mergers has important consequences to the distribution of BH masses and spins from this channel. We discuss these below. 

\begin{figure*}
   \centering  
   \includegraphics[width=0.95\textwidth]{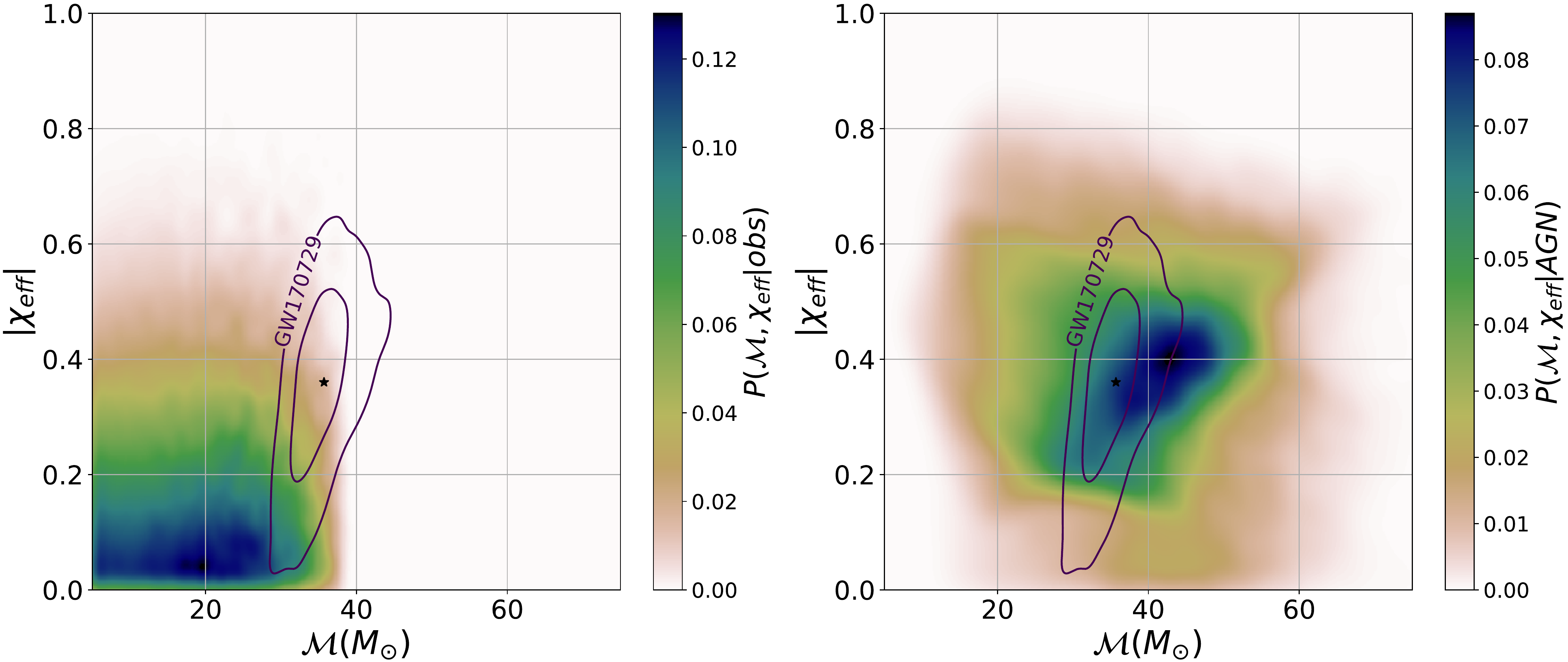}
   \caption{2D probability densities of the chirp mass $\mathcal{M}$ and effective spin parameter $\chi_{\rm eff}$ for BBHs detected by LIGO/Virgo. For the distributions we used the BBHs detected during LIGO/Virgo's O1 and O2 runs other than GW170729 (left) and for second and higher-generation mergers in the AGN model presented here (right). Also shown on both sides are the reconstructed parameters of GW170729, for its most likely values, 50\% and 90\% confidence regions.}
\label{fig:KDEAGN}
\end{figure*}


{\it Mass distribution.---} In Fig. \ref{fig:mass} we show the distribution of BH masses for different generations of mergers. For 1g mergers, we define $m_1>m_2$, while for $n$g mergers ($n>1$), $m_1$ is the mass of the BH from generation $n$. We see that the mass distribution increases significantly for higher-generation mergers, as expected. We find that about 30\% of mergers will have a BH with mass higher than the 50\,M$_\odot$ upper limit expected from stellar evolution \cite{2017ApJ...836..244W,2016A&A...594A..97B,2018MNRAS.480.2011G}. We account for detectability with LIGO/Virgo that favors heavier BHs using an estimated detection volume as a function of the binary's chirp mass $\mathcal{M}\equiv (m_1m_2)^{3/5}(m_1+m_2)^{-1/5}$. We find that 40\% of the detected mergers will have $m_1 > 50\,$M$_\odot$.

The mass distribution is virtually unaffected by whether higher-generation BH spins are aligned or anti-aligned with the binary orbit.


{\it Could both black holes be higher generation?.---} Since the merger remnant is expected to stay close to the migration trap, and since BHs move to the trap faster than the characteristic frequency of the AGN-alignment of new BHs, we expect one BH to "collect" all new incoming BHs. Therefore, one of the BHs in the binary should always be 1g. One exception is if a previously formed binary enters the disk and merges due to dynamical friction before reaching the trap. For simplicity we don't consider this case here due to the uncertain fraction of binaries in galactic nuclei. Using numerical simulations we estimate that if 10\% of the BHs in galactic nuclei reside in binaries, then in about 5\% of the mergers will both BHs be higher generation.


{\it Spin distribution.---} We derived the distribution of the binaries' effective spin $\chi_{\rm eff}$. We show its distribution in Fig. \ref{fig:spin} for different generations of mergers assuming that 50\% of the binaries have orbital angular momentum aligned with the AGN disk, and 50\% anti-aligned. We see the prominent peak around $\chi_{\rm eff}\sim0.4$. \addb{This is due to the typical mass ratio of 2g binaries. For an equal-mass 1g merger, the resulting BH has $\approx2$ times the mass of the initial BHs, and a spin of $\sim0.7$ \cite{2008ApJ...684..822B}. A binary of this BH and a spinless BH as massive as the original BHs has $\chi_{\rm eff}\sim2\times0.7 / (2+1)\approx0.45$. A similar argument can be applied to explain the two peaks in the 3g case in Fig. \ref{fig:spin}.} Otherwise we see that the spin distribution is broad. We find that this broad distribution is qualitatively similar for mergers with alignment and anti-alignment fractions other than 50\% as well, with the exception that anti-aligned spins lead to negative $\chi_{\rm eff}$ values\footnote{Hydrodynamical simulations by Secunda et al. (in prep) indicate aligned:anti-aligned ratios are likely in the 1:1 to 1:2 range.}.


{\it The case of GW170729.---} The BBH merger GW170729 has the largest mass, $m_1=50.6^{+16.6}_{-10.2}$\,M$_\odot$, and the largest measured spin, $\chi_{\rm eff}=0.36^{+0.21}_{-0.25}$, among all detected gravitational wave events \cite{2018arXiv181112907T}, making it particularly interesting to examine as a potential candidate from the AGN channel.

To establish whether GW170729 occurred in an AGN disk, we followed the method of \cite{2019arXiv190307813K}. We compared the merger's reconstructed parameters to the model presented here as well as to the parameters of the other BBH mergers LIGO/Virgo have detected so far \cite{2018arXiv181112907T}. 

\cite{2018arXiv181112940T} finds that the distribution of the masses of the 9 detected BBH mergers, excluding GW170729, can be approximated by a power-law distribution for the heavier mass, $m_{1}^{-\alpha}$ with $\alpha=1.6$, and a BH mass range of $[5\,\mbox{M}_\odot,45\,\mbox{M}_\odot]$, along with a uniform mass ratio distribution within $5\,\mbox{M}_\odot<m_2<m_1$. They find that the spin distribution (excluding GW170729) is consistent with isotropic directional and approximately flat amplitude distribution within $a\in[0,0.8]$. We adopted this model as our null hypothesis. 

For the signal hypothesis we adopted the joint mass and spin distribution of our AGN model. We characterize this distribution with the binary's chirp mass $\mathcal{M}$ and effective inspiral spin $\chi_{\rm eff}$. 

For both hypotheses we weight the signal- and null-hypothesis probability densities with the volume within which binary mergers with the given parameter can be detected. 

Our obtained distributions for $\mathcal{M}$ and $|\chi_{\rm eff}|$ for the observed and hierarchical-AGN cases are shown in Fig. \ref{fig:KDEAGN}. We see that second or higher-generation mergers in the AGN channel generally produce similar $\mathcal{M}$ and $\chi_{\rm eff}$ as observed for GW170729. Values from the 9 LIGO/Virgo observations are typically lower. First-generation mergers in AGNs result in comparable distribution as the LIGO/Virgo events.  

Nevertheless, there is not sufficient statistical evidence to confidently determine the formation channels for this event. We calculated the Bayesian odds ratio $P($AGN$\mid$GW170729$)/P($obs.$\mid$GW170729$)$. While the parameters of GW170729 are 5 times more likely to arise from our hierarchical-AGN distribution than from that of the null-hypothesis, taking into account a prior probability ratio $P(\mbox{AGN})/P(\mbox{obs/})=0.1-0.4$ \cite{2019ApJ...876..122Y}, we find that the odds ratio is $\sim 1$. More, similar events will be needed to probe this channel with high significance. In addition, other hierarchical-merger models could also explain GW170729 than the 9 LIGO/Virgo observations \cite{2019arXiv190307813K,2019arXiv190306742C}, although results at this point are also inconclusive (odds ratios are $\lesssim 3$).


{\it Conclusion.---} We examined the prevalence and expected mass/spin parameters of hierarchical mergers in AGN disks. Our conclusions are the following:
\begin{itemize}
\item Hierarchical mergers are the norm rather than the exception in the migration traps of AGN disks, \add{if these traps exist}. As BHs accumulate in the migration trap they merge with the BHs already there, resulting in a chain of consecutive mergers. For our fiducial parameters over 50\% of BH mergers are higher-generation.
\item Hierarchical mergers result in heavy BHs. In about 40\% of the detected mergers, one of the BHs is heavier than 50\,M$_\odot$.
\item Hierarchical mergers in AGN disks will naturally lead to aligned spins with the AGN disk. This leads to aligned or anti-aligned spins with the binary orbit. A broad range of spins are possible from about $0.2-0.9$. In particular, anti-aligned spins are a unique possibility in this model compared to other channels \cite{2018MNRAS.480L..58A}. \add{Spin alignment makes this channel observationally distinguishable from other hierarchical merger channels \cite{2017PhRvD..95l4046G,2017ApJ...840L..24F,2019arXiv190307813K}.}
\item Finding high-mass, non-zero spin BH mergers in AGNs will also probe the physics of orbital alignment and the development of migration traps.
\item We find that the heaviest BBH merger detected so far, GW170729, has similar $\mathcal{M}$ and $\chi_{\rm eff}$ to those expected from second or higher-generation mergers in AGNs (see Fig. \ref{fig:KDEAGN}). Nevertheless, there is not sufficient statistical evidence to differentiate between an AGN origin and the same channel as the other 9 events detected by LIGO/Virgo so far.
\end{itemize}


\begin{acknowledgments}
The authors thank Christopher Berry, Emanuele Berti, Thomas Dent, Davide Gerosa and Brian Metzger for their useful suggestions. YY and IB are grateful to the University of Florida for their generous support. VG acknowledges Inspire division, DST, Government of India for the fellowship support. ZH acknowledges support from NSF grant 1715661 and NASA grants NNX17AL82G and NNX15AB19G. This project has received funding from the European Research Council (ERC) under the European Union’s Horizon 2020 research and innovation programme under grant agreement No 638435 (GalNUC) and by the Hungarian National Research, Development, and Innovation Office grant NKFIH KH-125675 (to BK). SK acknowledges the generous support of the NSF under grant number PHY-1806165. SM and ZM thank Columbia University in the City of New York for their generous support. The Columbia Experimental Gravity group is grateful for the generous support of the National Science Foundation under grant PHY-1708028. KESF and BM are supported by NSF grant 1831412. ROS is supported by NSF PHY 1707965.
\end{acknowledgments}

\bibliographystyle{apsrev4-2}
\bibliography{Refs}

\begin{thebibliography}{63}%
\makeatletter
\providecommand \@ifxundefined [1]{%
 \@ifx{#1\undefined}
}%
\providecommand \@ifnum [1]{%
 \ifnum #1\expandafter \@firstoftwo
 \else \expandafter \@secondoftwo
 \fi
}%
\providecommand \@ifx [1]{%
 \ifx #1\expandafter \@firstoftwo
 \else \expandafter \@secondoftwo
 \fi
}%
\providecommand \natexlab [1]{#1}%
\providecommand \enquote  [1]{``#1''}%
\providecommand \bibnamefont  [1]{#1}%
\providecommand \bibfnamefont [1]{#1}%
\providecommand \citenamefont [1]{#1}%
\providecommand \href@noop [0]{\@secondoftwo}%
\providecommand \href [0]{\begingroup \@sanitize@url \@href}%
\providecommand \@href[1]{\@@startlink{#1}\@@href}%
\providecommand \@@href[1]{\endgroup#1\@@endlink}%
\providecommand \@sanitize@url [0]{\catcode `\\12\catcode `\$12\catcode
  `\&12\catcode `\#12\catcode `\^12\catcode `\_12\catcode `\%12\relax}%
\providecommand \@@startlink[1]{}%
\providecommand \@@endlink[0]{}%
\providecommand \url  [0]{\begingroup\@sanitize@url \@url }%
\providecommand \@url [1]{\endgroup\@href {#1}{\urlprefix }}%
\providecommand \urlprefix  [0]{URL }%
\providecommand \Eprint [0]{\href }%
\providecommand \doibase [0]{https://doi.org/}%
\providecommand \selectlanguage [0]{\@gobble}%
\providecommand \bibinfo  [0]{\@secondoftwo}%
\providecommand \bibfield  [0]{\@secondoftwo}%
\providecommand \translation [1]{[#1]}%
\providecommand \BibitemOpen [0]{}%
\providecommand \bibitemStop [0]{}%
\providecommand \bibitemNoStop [0]{.\EOS\space}%
\providecommand \EOS [0]{\spacefactor3000\relax}%
\providecommand \BibitemShut  [1]{\csname bibitem#1\endcsname}%
\let\auto@bib@innerbib\@empty
\bibitem [{\citenamefont {{Aasi}}\ \emph {et~al.}(2015)\citenamefont {{Aasi}}
  \emph {et~al.}}]{2015CQGra..32g4001L}%
  \BibitemOpen
  \bibfield  {author} {\bibinfo {author} {\bibfnamefont {J.}~\bibnamefont
  {{Aasi}}} \emph {et~al.},\ }\href
  {https://doi.org/10.1088/0264-9381/32/7/074001} {\bibfield  {journal}
  {\bibinfo  {journal} {\cqg}\ }\textbf {\bibinfo {volume} {32}},\ \bibinfo
  {eid} {074001} (\bibinfo {year} {2015})}\BibitemShut {NoStop}%
\bibitem [{\citenamefont {{Acernese}}\ \emph {et~al.}(2015)\citenamefont
  {{Acernese}} \emph {et~al.}}]{2015CQGra..32b4001A}%
  \BibitemOpen
  \bibfield  {author} {\bibinfo {author} {\bibfnamefont {F.}~\bibnamefont
  {{Acernese}}} \emph {et~al.},\ }\href
  {https://doi.org/10.1088/0264-9381/32/2/024001} {\bibfield  {journal}
  {\bibinfo  {journal} {\cqg}\ }\textbf {\bibinfo {volume} {32}},\ \bibinfo
  {eid} {024001} (\bibinfo {year} {2015})}\BibitemShut {NoStop}%
\bibitem [{\citenamefont {{Abbott}}\ \emph
  {et~al.}(2019{\natexlab{a}})\citenamefont {{Abbott}} \emph
  {et~al.}}]{2018arXiv181112907T}%
  \BibitemOpen
  \bibfield  {author} {\bibinfo {author} {\bibfnamefont {B.~P.}\ \bibnamefont
  {{Abbott}}} \emph {et~al.},\ }\href
  {https://doi.org/10.1103/PhysRevX.9.031040} {\bibfield  {journal} {\bibinfo
  {journal} {Phys. Rev. X}\ }\textbf {\bibinfo {volume} {9}},\ \bibinfo {pages}
  {031040} (\bibinfo {year} {2019}{\natexlab{a}})}\BibitemShut {NoStop}%
\bibitem [{\citenamefont {{Venumadhav}}\ \emph {et~al.}(2019)\citenamefont
  {{Venumadhav}}, \citenamefont {{Zackay}}, \citenamefont {{Roulet}},
  \citenamefont {{Dai}},\ and\ \citenamefont
  {{Zaldarriaga}}}]{2019arXiv190407214V}%
  \BibitemOpen
  \bibfield  {author} {\bibinfo {author} {\bibfnamefont {T.}~\bibnamefont
  {{Venumadhav}}}, \bibinfo {author} {\bibfnamefont {B.}~\bibnamefont
  {{Zackay}}}, \bibinfo {author} {\bibfnamefont {J.}~\bibnamefont {{Roulet}}},
  \bibinfo {author} {\bibfnamefont {L.}~\bibnamefont {{Dai}}},\ and\ \bibinfo
  {author} {\bibfnamefont {M.}~\bibnamefont {{Zaldarriaga}}},\ }\href@noop {}
  {\bibfield  {journal} {\bibinfo  {journal} {arXiv:1904.07214}\ } (\bibinfo
  {year} {2019})}\BibitemShut {NoStop}%
\bibitem [{\citenamefont {{Abbott}}\ \emph {et~al.}(2018)\citenamefont
  {{Abbott}} \emph {et~al.}}]{2018LRR....21....3A}%
  \BibitemOpen
  \bibfield  {author} {\bibinfo {author} {\bibfnamefont {B.~P.}\ \bibnamefont
  {{Abbott}}} \emph {et~al.},\ }\href
  {https://doi.org/10.1007/s41114-018-0012-9} {\bibfield  {journal} {\bibinfo
  {journal} {Living Rev. Relativ.}\ }\textbf {\bibinfo {volume} {21}},\
  \bibinfo {eid} {3} (\bibinfo {year} {2018})}\BibitemShut {NoStop}%
\bibitem [{\citenamefont {{Belczynski}}\ \emph {et~al.}(2010)\citenamefont
  {{Belczynski}}, \citenamefont {{Bulik}}, \citenamefont {{Fryer}},
  \citenamefont {{Ruiter}}, \citenamefont {{Valsecchi}}, \citenamefont
  {{Vink}},\ and\ \citenamefont {{Hurley}}}]{2010ApJ...714.1217B}%
  \BibitemOpen
  \bibfield  {author} {\bibinfo {author} {\bibfnamefont {K.}~\bibnamefont
  {{Belczynski}}}, \bibinfo {author} {\bibfnamefont {T.}~\bibnamefont
  {{Bulik}}}, \bibinfo {author} {\bibfnamefont {C.~L.}\ \bibnamefont
  {{Fryer}}}, \bibinfo {author} {\bibfnamefont {A.}~\bibnamefont {{Ruiter}}},
  \bibinfo {author} {\bibfnamefont {F.}~\bibnamefont {{Valsecchi}}}, \bibinfo
  {author} {\bibfnamefont {J.~S.}\ \bibnamefont {{Vink}}},\ and\ \bibinfo
  {author} {\bibfnamefont {J.~R.}\ \bibnamefont {{Hurley}}},\ }\href
  {https://doi.org/10.1088/0004-637X/714/2/1217} {\bibfield  {journal}
  {\bibinfo  {journal} {\apj}\ }\textbf {\bibinfo {volume} {714}},\ \bibinfo
  {pages} {1217} (\bibinfo {year} {2010})}\BibitemShut {NoStop}%
\bibitem [{\citenamefont {Postnov}\ and\ \citenamefont
  {Yungelson}(2014)}]{Postnov2014}%
  \BibitemOpen
  \bibfield  {author} {\bibinfo {author} {\bibfnamefont {K.~A.}\ \bibnamefont
  {Postnov}}\ and\ \bibinfo {author} {\bibfnamefont {L.~R.}\ \bibnamefont
  {Yungelson}},\ }\href {https://doi.org/10.12942/lrr-2014-3} {\bibfield
  {journal} {\bibinfo  {journal} {Living Rev. Relativ.}\ }\textbf {\bibinfo
  {volume} {17}},\ \bibinfo {pages} {3} (\bibinfo {year} {2014})}\BibitemShut
  {NoStop}%
\bibitem [{\citenamefont {{de Mink}}\ and\ \citenamefont
  {{Mandel}}(2016)}]{2016MNRAS.460.3545D}%
  \BibitemOpen
  \bibfield  {author} {\bibinfo {author} {\bibfnamefont {S.~E.}\ \bibnamefont
  {{de Mink}}}\ and\ \bibinfo {author} {\bibfnamefont {I.}~\bibnamefont
  {{Mandel}}},\ }\href {https://doi.org/10.1093/mnras/stw1219} {\bibfield
  {journal} {\bibinfo  {journal} {\mnras}\ }\textbf {\bibinfo {volume} {460}},\
  \bibinfo {pages} {3545} (\bibinfo {year} {2016})}\BibitemShut {NoStop}%
\bibitem [{\citenamefont {{Hopman}}\ and\ \citenamefont
  {{Alexander}}(2006)}]{2006ApJ...645L.133H}%
  \BibitemOpen
  \bibfield  {author} {\bibinfo {author} {\bibfnamefont {C.}~\bibnamefont
  {{Hopman}}}\ and\ \bibinfo {author} {\bibfnamefont {T.}~\bibnamefont
  {{Alexander}}},\ }\href {https://doi.org/10.1086/506273} {\bibfield
  {journal} {\bibinfo  {journal} {\apj}\ }\textbf {\bibinfo {volume} {645}},\
  \bibinfo {pages} {L133} (\bibinfo {year} {2006})}\BibitemShut {NoStop}%
\bibitem [{\citenamefont {{O'Leary}}\ \emph {et~al.}(2009)\citenamefont
  {{O'Leary}}, \citenamefont {{Kocsis}},\ and\ \citenamefont
  {{Loeb}}}]{2009MNRAS.395.2127O}%
  \BibitemOpen
  \bibfield  {author} {\bibinfo {author} {\bibfnamefont {R.~M.}\ \bibnamefont
  {{O'Leary}}}, \bibinfo {author} {\bibfnamefont {B.}~\bibnamefont
  {{Kocsis}}},\ and\ \bibinfo {author} {\bibfnamefont {A.}~\bibnamefont
  {{Loeb}}},\ }\href {https://doi.org/10.1111/j.1365-2966.2009.14653.x}
  {\bibfield  {journal} {\bibinfo  {journal} {\mnras}\ }\textbf {\bibinfo
  {volume} {395}},\ \bibinfo {pages} {2127} (\bibinfo {year}
  {2009})}\BibitemShut {NoStop}%
\bibitem [{\citenamefont {Benacquista}\ and\ \citenamefont
  {Downing}(2013)}]{Benacquista2013}%
  \BibitemOpen
  \bibfield  {author} {\bibinfo {author} {\bibfnamefont {M.~J.}\ \bibnamefont
  {Benacquista}}\ and\ \bibinfo {author} {\bibfnamefont {J.~M.~B.}\
  \bibnamefont {Downing}},\ }\href {https://doi.org/10.12942/lrr-2013-4}
  {\bibfield  {journal} {\bibinfo  {journal} {Living Rev. Relativ.}\ }\textbf
  {\bibinfo {volume} {16}},\ \bibinfo {pages} {4} (\bibinfo {year}
  {2013})}\BibitemShut {NoStop}%
\bibitem [{\citenamefont {{Breen}}\ and\ \citenamefont
  {{Heggie}}(2013)}]{2013MNRAS.432.2779B}%
  \BibitemOpen
  \bibfield  {author} {\bibinfo {author} {\bibfnamefont {P.~G.}\ \bibnamefont
  {{Breen}}}\ and\ \bibinfo {author} {\bibfnamefont {D.~C.}\ \bibnamefont
  {{Heggie}}},\ }\href {https://doi.org/10.1093/mnras/stt628} {\bibfield
  {journal} {\bibinfo  {journal} {\mnras}\ }\textbf {\bibinfo {volume} {432}},\
  \bibinfo {pages} {2779} (\bibinfo {year} {2013})}\BibitemShut {NoStop}%
\bibitem [{\citenamefont {{Rodriguez}}\ \emph
  {et~al.}(2016{\natexlab{a}})\citenamefont {{Rodriguez}}, \citenamefont
  {{Chatterjee}},\ and\ \citenamefont {{Rasio}}}]{2016PhRvD..93h4029R}%
  \BibitemOpen
  \bibfield  {author} {\bibinfo {author} {\bibfnamefont {C.~L.}\ \bibnamefont
  {{Rodriguez}}}, \bibinfo {author} {\bibfnamefont {S.}~\bibnamefont
  {{Chatterjee}}},\ and\ \bibinfo {author} {\bibfnamefont {F.~A.}\ \bibnamefont
  {{Rasio}}},\ }\href {https://doi.org/10.1103/PhysRevD.93.084029} {\bibfield
  {journal} {\bibinfo  {journal} {\prd}\ }\textbf {\bibinfo {volume} {93}},\
  \bibinfo {eid} {084029} (\bibinfo {year} {2016}{\natexlab{a}})}\BibitemShut
  {NoStop}%
\bibitem [{\citenamefont {{Rodriguez}}\ \emph
  {et~al.}(2016{\natexlab{b}})\citenamefont {{Rodriguez}}, \citenamefont
  {{Haster}}, \citenamefont {{Chatterjee}}, \citenamefont {{Kalogera}},\ and\
  \citenamefont {{Rasio}}}]{2016ApJ...824L...8R}%
  \BibitemOpen
  \bibfield  {author} {\bibinfo {author} {\bibfnamefont {C.~L.}\ \bibnamefont
  {{Rodriguez}}}, \bibinfo {author} {\bibfnamefont {C.-J.}\ \bibnamefont
  {{Haster}}}, \bibinfo {author} {\bibfnamefont {S.}~\bibnamefont
  {{Chatterjee}}}, \bibinfo {author} {\bibfnamefont {V.}~\bibnamefont
  {{Kalogera}}},\ and\ \bibinfo {author} {\bibfnamefont {F.~A.}\ \bibnamefont
  {{Rasio}}},\ }\href {https://doi.org/10.3847/2041-8205/824/1/L8} {\bibfield
  {journal} {\bibinfo  {journal} {\apj}\ }\textbf {\bibinfo {volume} {824}},\
  \bibinfo {eid} {L8} (\bibinfo {year} {2016}{\natexlab{b}})}\BibitemShut
  {NoStop}%
\bibitem [{\citenamefont {{Rodriguez}}\ \emph {et~al.}(2018)\citenamefont
  {{Rodriguez}}, \citenamefont {{Amaro-Seoane}}, \citenamefont {{Chatterjee}},\
  and\ \citenamefont {{Rasio}}}]{2018PhRvL.120o1101R}%
  \BibitemOpen
  \bibfield  {author} {\bibinfo {author} {\bibfnamefont {C.~L.}\ \bibnamefont
  {{Rodriguez}}}, \bibinfo {author} {\bibfnamefont {P.}~\bibnamefont
  {{Amaro-Seoane}}}, \bibinfo {author} {\bibfnamefont {S.}~\bibnamefont
  {{Chatterjee}}},\ and\ \bibinfo {author} {\bibfnamefont {F.~A.}\ \bibnamefont
  {{Rasio}}},\ }\href {https://doi.org/10.1103/PhysRevLett.120.151101}
  {\bibfield  {journal} {\bibinfo  {journal} {\prl}\ }\textbf {\bibinfo
  {volume} {120}},\ \bibinfo {eid} {151101} (\bibinfo {year}
  {2018})}\BibitemShut {NoStop}%
\bibitem [{\citenamefont {{Antonini}}\ and\ \citenamefont
  {{Rasio}}(2016)}]{2016ApJ...831..187A}%
  \BibitemOpen
  \bibfield  {author} {\bibinfo {author} {\bibfnamefont {F.}~\bibnamefont
  {{Antonini}}}\ and\ \bibinfo {author} {\bibfnamefont {F.~A.}\ \bibnamefont
  {{Rasio}}},\ }\href {https://doi.org/10.3847/0004-637X/831/2/187} {\bibfield
  {journal} {\bibinfo  {journal} {\apj}\ }\textbf {\bibinfo {volume} {831}},\
  \bibinfo {eid} {187} (\bibinfo {year} {2016})}\BibitemShut {NoStop}%
\bibitem [{\citenamefont {{Wang}}\ \emph {et~al.}(2016)\citenamefont {{Wang}},
  \citenamefont {{Spurzem}}, \citenamefont {{Aarseth}}, \citenamefont
  {{Giersz}}, \citenamefont {{Askar}}, \citenamefont {{Berczik}}, \citenamefont
  {{Naab}}, \citenamefont {{Schadow}},\ and\ \citenamefont
  {{Kouwenhoven}}}]{2016MNRAS.458.1450W}%
  \BibitemOpen
  \bibfield  {author} {\bibinfo {author} {\bibfnamefont {L.}~\bibnamefont
  {{Wang}}}, \bibinfo {author} {\bibfnamefont {R.}~\bibnamefont {{Spurzem}}},
  \bibinfo {author} {\bibfnamefont {S.}~\bibnamefont {{Aarseth}}}, \bibinfo
  {author} {\bibfnamefont {M.}~\bibnamefont {{Giersz}}}, \bibinfo {author}
  {\bibfnamefont {A.}~\bibnamefont {{Askar}}}, \bibinfo {author} {\bibfnamefont
  {P.}~\bibnamefont {{Berczik}}}, \bibinfo {author} {\bibfnamefont
  {T.}~\bibnamefont {{Naab}}}, \bibinfo {author} {\bibfnamefont
  {R.}~\bibnamefont {{Schadow}}},\ and\ \bibinfo {author} {\bibfnamefont
  {M.~B.~N.}\ \bibnamefont {{Kouwenhoven}}},\ }\href
  {https://doi.org/10.1093/mnras/stw274} {\bibfield  {journal} {\bibinfo
  {journal} {\mnras}\ }\textbf {\bibinfo {volume} {458}},\ \bibinfo {pages}
  {1450} (\bibinfo {year} {2016})}\BibitemShut {NoStop}%
\bibitem [{\citenamefont {{Morris}}(1993)}]{1993ApJ...408..496M}%
  \BibitemOpen
  \bibfield  {author} {\bibinfo {author} {\bibfnamefont {M.}~\bibnamefont
  {{Morris}}},\ }\href {https://doi.org/10.1086/172607} {\bibfield  {journal}
  {\bibinfo  {journal} {\apj}\ }\textbf {\bibinfo {volume} {408}},\ \bibinfo
  {pages} {496} (\bibinfo {year} {1993})}\BibitemShut {NoStop}%
\bibitem [{\citenamefont {{Miralda-Escud{\'e}}}\ and\ \citenamefont
  {{Gould}}(2000)}]{2000ApJ...545..847M}%
  \BibitemOpen
  \bibfield  {author} {\bibinfo {author} {\bibfnamefont {J.}~\bibnamefont
  {{Miralda-Escud{\'e}}}}\ and\ \bibinfo {author} {\bibfnamefont
  {A.}~\bibnamefont {{Gould}}},\ }\href {https://doi.org/10.1086/317837}
  {\bibfield  {journal} {\bibinfo  {journal} {\apj}\ }\textbf {\bibinfo
  {volume} {545}},\ \bibinfo {pages} {847} (\bibinfo {year}
  {2000})}\BibitemShut {NoStop}%
\bibitem [{\citenamefont {{Antonini}}(2014)}]{2014ApJ...794..106A}%
  \BibitemOpen
  \bibfield  {author} {\bibinfo {author} {\bibfnamefont {F.}~\bibnamefont
  {{Antonini}}},\ }\href {https://doi.org/10.1088/0004-637X/794/2/106}
  {\bibfield  {journal} {\bibinfo  {journal} {\apj}\ }\textbf {\bibinfo
  {volume} {794}},\ \bibinfo {eid} {106} (\bibinfo {year} {2014})}\BibitemShut
  {NoStop}%
\bibitem [{\citenamefont {{Hailey}}\ \emph {et~al.}(2018)\citenamefont
  {{Hailey}}, \citenamefont {{Mori}}, \citenamefont {{Bauer}}, \citenamefont
  {{Berkowitz}}, \citenamefont {{Hong}},\ and\ \citenamefont
  {{Hord}}}]{2018Natur.556...70H}%
  \BibitemOpen
  \bibfield  {author} {\bibinfo {author} {\bibfnamefont {C.~J.}\ \bibnamefont
  {{Hailey}}}, \bibinfo {author} {\bibfnamefont {K.}~\bibnamefont {{Mori}}},
  \bibinfo {author} {\bibfnamefont {F.~E.}\ \bibnamefont {{Bauer}}}, \bibinfo
  {author} {\bibfnamefont {M.~E.}\ \bibnamefont {{Berkowitz}}}, \bibinfo
  {author} {\bibfnamefont {J.}~\bibnamefont {{Hong}}},\ and\ \bibinfo {author}
  {\bibfnamefont {B.~J.}\ \bibnamefont {{Hord}}},\ }\href
  {https://doi.org/10.1038/nature25029} {\bibfield  {journal} {\bibinfo
  {journal} {\nat}\ }\textbf {\bibinfo {volume} {556}},\ \bibinfo {pages} {70}
  (\bibinfo {year} {2018})}\BibitemShut {NoStop}%
\bibitem [{\citenamefont {{Generozov}}\ \emph {et~al.}(2018)\citenamefont
  {{Generozov}}, \citenamefont {{Stone}}, \citenamefont {{Metzger}},\ and\
  \citenamefont {{Ostriker}}}]{2018MNRAS.478.4030G}%
  \BibitemOpen
  \bibfield  {author} {\bibinfo {author} {\bibfnamefont {A.}~\bibnamefont
  {{Generozov}}}, \bibinfo {author} {\bibfnamefont {N.~C.}\ \bibnamefont
  {{Stone}}}, \bibinfo {author} {\bibfnamefont {B.~D.}\ \bibnamefont
  {{Metzger}}},\ and\ \bibinfo {author} {\bibfnamefont {J.~P.}\ \bibnamefont
  {{Ostriker}}},\ }\href {https://doi.org/10.1093/mnras/sty1262} {\bibfield
  {journal} {\bibinfo  {journal} {\mnras}\ }\textbf {\bibinfo {volume} {478}},\
  \bibinfo {pages} {4030} (\bibinfo {year} {2018})}\BibitemShut {NoStop}%
\bibitem [{\citenamefont {{Syer}}\ \emph {et~al.}(1991)\citenamefont {{Syer}},
  \citenamefont {{Clarke}},\ and\ \citenamefont
  {{Rees}}}]{1991MNRAS.250..505S}%
  \BibitemOpen
  \bibfield  {author} {\bibinfo {author} {\bibfnamefont {D.}~\bibnamefont
  {{Syer}}}, \bibinfo {author} {\bibfnamefont {C.~J.}\ \bibnamefont
  {{Clarke}}},\ and\ \bibinfo {author} {\bibfnamefont {M.~J.}\ \bibnamefont
  {{Rees}}},\ }\href {https://doi.org/10.1093/mnras/250.3.505} {\bibfield
  {journal} {\bibinfo  {journal} {\mnras}\ }\textbf {\bibinfo {volume} {250}},\
  \bibinfo {pages} {505} (\bibinfo {year} {1991})}\BibitemShut {NoStop}%
\bibitem [{\citenamefont {{Artymowicz}}\ \emph {et~al.}(1993)\citenamefont
  {{Artymowicz}}, \citenamefont {{Lin}},\ and\ \citenamefont
  {{Wampler}}}]{1993ApJ...409..592A}%
  \BibitemOpen
  \bibfield  {author} {\bibinfo {author} {\bibfnamefont {P.}~\bibnamefont
  {{Artymowicz}}}, \bibinfo {author} {\bibfnamefont {D.~N.~C.}\ \bibnamefont
  {{Lin}}},\ and\ \bibinfo {author} {\bibfnamefont {E.~J.}\ \bibnamefont
  {{Wampler}}},\ }\href {https://doi.org/10.1086/172690} {\bibfield  {journal}
  {\bibinfo  {journal} {\apj}\ }\textbf {\bibinfo {volume} {409}},\ \bibinfo
  {pages} {592} (\bibinfo {year} {1993})}\BibitemShut {NoStop}%
\bibitem [{\citenamefont {{Goodman}}\ and\ \citenamefont
  {{Tan}}(2004)}]{2004ApJ...608..108G}%
  \BibitemOpen
  \bibfield  {author} {\bibinfo {author} {\bibfnamefont {J.}~\bibnamefont
  {{Goodman}}}\ and\ \bibinfo {author} {\bibfnamefont {J.~C.}\ \bibnamefont
  {{Tan}}},\ }\href {https://doi.org/10.1086/386360} {\bibfield  {journal}
  {\bibinfo  {journal} {\apj}\ }\textbf {\bibinfo {volume} {608}},\ \bibinfo
  {pages} {108} (\bibinfo {year} {2004})}\BibitemShut {NoStop}%
\bibitem [{\citenamefont {{McKernan}}\ \emph {et~al.}(2012)\citenamefont
  {{McKernan}}, \citenamefont {{Ford}}, \citenamefont {{Lyra}},\ and\
  \citenamefont {{Perets}}}]{2012MNRAS.425..460M}%
  \BibitemOpen
  \bibfield  {author} {\bibinfo {author} {\bibfnamefont {B.}~\bibnamefont
  {{McKernan}}}, \bibinfo {author} {\bibfnamefont {K.~E.~S.}\ \bibnamefont
  {{Ford}}}, \bibinfo {author} {\bibfnamefont {W.}~\bibnamefont {{Lyra}}},\
  and\ \bibinfo {author} {\bibfnamefont {H.~B.}\ \bibnamefont {{Perets}}},\
  }\href {https://doi.org/10.1111/j.1365-2966.2012.21486.x} {\bibfield
  {journal} {\bibinfo  {journal} {\mnras}\ }\textbf {\bibinfo {volume} {425}},\
  \bibinfo {pages} {460} (\bibinfo {year} {2012})}\BibitemShut {NoStop}%
\bibitem [{\citenamefont {{McKernan}}\ \emph {et~al.}(2014)\citenamefont
  {{McKernan}}, \citenamefont {{Ford}}, \citenamefont {{Kocsis}}, \citenamefont
  {{Lyra}},\ and\ \citenamefont {{Winter}}}]{2014MNRAS.441..900M}%
  \BibitemOpen
  \bibfield  {author} {\bibinfo {author} {\bibfnamefont {B.}~\bibnamefont
  {{McKernan}}}, \bibinfo {author} {\bibfnamefont {K.~E.~S.}\ \bibnamefont
  {{Ford}}}, \bibinfo {author} {\bibfnamefont {B.}~\bibnamefont {{Kocsis}}},
  \bibinfo {author} {\bibfnamefont {W.}~\bibnamefont {{Lyra}}},\ and\ \bibinfo
  {author} {\bibfnamefont {L.~M.}\ \bibnamefont {{Winter}}},\ }\href
  {https://doi.org/10.1093/mnras/stu553} {\bibfield  {journal} {\bibinfo
  {journal} {\mnras}\ }\textbf {\bibinfo {volume} {441}},\ \bibinfo {pages}
  {900} (\bibinfo {year} {2014})}\BibitemShut {NoStop}%
\bibitem [{\citenamefont {{Bartos}}\ \emph
  {et~al.}(2017{\natexlab{a}})\citenamefont {{Bartos}}, \citenamefont
  {{Kocsis}}, \citenamefont {{Haiman}},\ and\ \citenamefont
  {{M{\'a}rka}}}]{2017ApJ...835..165B}%
  \BibitemOpen
  \bibfield  {author} {\bibinfo {author} {\bibfnamefont {I.}~\bibnamefont
  {{Bartos}}}, \bibinfo {author} {\bibfnamefont {B.}~\bibnamefont {{Kocsis}}},
  \bibinfo {author} {\bibfnamefont {Z.}~\bibnamefont {{Haiman}}},\ and\
  \bibinfo {author} {\bibfnamefont {S.}~\bibnamefont {{M{\'a}rka}}},\ }\href
  {https://doi.org/10.3847/1538-4357/835/2/165} {\bibfield  {journal} {\bibinfo
   {journal} {\apj}\ }\textbf {\bibinfo {volume} {835}},\ \bibinfo {eid} {165}
  (\bibinfo {year} {2017}{\natexlab{a}})}\BibitemShut {NoStop}%
\bibitem [{\citenamefont {{Levin}}(2007)}]{2007MNRAS.374..515L}%
  \BibitemOpen
  \bibfield  {author} {\bibinfo {author} {\bibfnamefont {Y.}~\bibnamefont
  {{Levin}}},\ }\href {https://doi.org/10.1111/j.1365-2966.2006.11155.x}
  {\bibfield  {journal} {\bibinfo  {journal} {\mnras}\ }\textbf {\bibinfo
  {volume} {374}},\ \bibinfo {pages} {515} (\bibinfo {year}
  {2007})}\BibitemShut {NoStop}%
\bibitem [{\citenamefont {{Stone}}\ \emph {et~al.}(2017)\citenamefont
  {{Stone}}, \citenamefont {{Metzger}},\ and\ \citenamefont
  {{Haiman}}}]{2017MNRAS.464..946S}%
  \BibitemOpen
  \bibfield  {author} {\bibinfo {author} {\bibfnamefont {N.~C.}\ \bibnamefont
  {{Stone}}}, \bibinfo {author} {\bibfnamefont {B.~D.}\ \bibnamefont
  {{Metzger}}},\ and\ \bibinfo {author} {\bibfnamefont {Z.}~\bibnamefont
  {{Haiman}}},\ }\href {https://doi.org/10.1093/mnras/stw2260} {\bibfield
  {journal} {\bibinfo  {journal} {\mnras}\ }\textbf {\bibinfo {volume} {464}},\
  \bibinfo {pages} {946} (\bibinfo {year} {2017})}\BibitemShut {NoStop}%
\bibitem [{\citenamefont {{Bellovary}}\ \emph {et~al.}(2016)\citenamefont
  {{Bellovary}}, \citenamefont {{Mac Low}}, \citenamefont {{McKernan}},\ and\
  \citenamefont {{Ford}}}]{2016ApJ...819L..17B}%
  \BibitemOpen
  \bibfield  {author} {\bibinfo {author} {\bibfnamefont {J.~M.}\ \bibnamefont
  {{Bellovary}}}, \bibinfo {author} {\bibfnamefont {M.-M.}\ \bibnamefont {{Mac
  Low}}}, \bibinfo {author} {\bibfnamefont {B.}~\bibnamefont {{McKernan}}},\
  and\ \bibinfo {author} {\bibfnamefont {K.~E.~S.}\ \bibnamefont {{Ford}}},\
  }\href {https://doi.org/10.3847/2041-8205/819/2/L17} {\bibfield  {journal}
  {\bibinfo  {journal} {\apjl}\ }\textbf {\bibinfo {volume} {819}},\ \bibinfo
  {eid} {L17} (\bibinfo {year} {2016})}\BibitemShut {NoStop}%
\bibitem [{\citenamefont {{Secunda}}\ \emph {et~al.}(2019)\citenamefont
  {{Secunda}}, \citenamefont {{Bellovary}}, \citenamefont {{Mac Low}},
  \citenamefont {{Ford}}, \citenamefont {{McKernan}}, \citenamefont {{Leigh}},
  \citenamefont {{Lyra}},\ and\ \citenamefont
  {{S{\'a}ndor}}}]{2018arXiv180702859S}%
  \BibitemOpen
  \bibfield  {author} {\bibinfo {author} {\bibfnamefont {A.}~\bibnamefont
  {{Secunda}}}, \bibinfo {author} {\bibfnamefont {J.}~\bibnamefont
  {{Bellovary}}}, \bibinfo {author} {\bibfnamefont {M.-M.}\ \bibnamefont {{Mac
  Low}}}, \bibinfo {author} {\bibfnamefont {K.~E.~S.}\ \bibnamefont {{Ford}}},
  \bibinfo {author} {\bibfnamefont {B.}~\bibnamefont {{McKernan}}}, \bibinfo
  {author} {\bibfnamefont {N.~W.~C.}\ \bibnamefont {{Leigh}}}, \bibinfo
  {author} {\bibfnamefont {W.}~\bibnamefont {{Lyra}}},\ and\ \bibinfo {author}
  {\bibfnamefont {Z.}~\bibnamefont {{S{\'a}ndor}}},\ }\href
  {https://doi.org/10.3847/1538-4357/ab20ca} {\bibfield  {journal} {\bibinfo
  {journal} {\apj}\ }\textbf {\bibinfo {volume} {878}},\ \bibinfo {eid} {85}
  (\bibinfo {year} {2019})}\BibitemShut {NoStop}%
\bibitem [{\citenamefont {{Kulkarni}}\ \emph {et~al.}(1993)\citenamefont
  {{Kulkarni}}, \citenamefont {{Hut}},\ and\ \citenamefont
  {{McMillan}}}]{1993Natur.364..421K}%
  \BibitemOpen
  \bibfield  {author} {\bibinfo {author} {\bibfnamefont {S.~R.}\ \bibnamefont
  {{Kulkarni}}}, \bibinfo {author} {\bibfnamefont {P.}~\bibnamefont {{Hut}}},\
  and\ \bibinfo {author} {\bibfnamefont {S.}~\bibnamefont {{McMillan}}},\
  }\href {https://doi.org/10.1038/364421a0} {\bibfield  {journal} {\bibinfo
  {journal} {\nat}\ }\textbf {\bibinfo {volume} {364}},\ \bibinfo {pages} {421}
  (\bibinfo {year} {1993})}\BibitemShut {NoStop}%
\bibitem [{\citenamefont {{Sigurdsson}}\ and\ \citenamefont
  {{Hernquist}}(1993)}]{1993Natur.364..423S}%
  \BibitemOpen
  \bibfield  {author} {\bibinfo {author} {\bibfnamefont {S.}~\bibnamefont
  {{Sigurdsson}}}\ and\ \bibinfo {author} {\bibfnamefont {L.}~\bibnamefont
  {{Hernquist}}},\ }\href {https://doi.org/10.1038/364423a0} {\bibfield
  {journal} {\bibinfo  {journal} {\nat}\ }\textbf {\bibinfo {volume} {364}},\
  \bibinfo {pages} {423} (\bibinfo {year} {1993})}\BibitemShut {NoStop}%
\bibitem [{\citenamefont {{Portegies Zwart}}\ and\ \citenamefont
  {{McMillan}}(2002)}]{2002ApJ...576..899P}%
  \BibitemOpen
  \bibfield  {author} {\bibinfo {author} {\bibfnamefont {S.~F.}\ \bibnamefont
  {{Portegies Zwart}}}\ and\ \bibinfo {author} {\bibfnamefont {S.~L.~W.}\
  \bibnamefont {{McMillan}}},\ }\href {https://doi.org/10.1086/341798}
  {\bibfield  {journal} {\bibinfo  {journal} {\apj}\ }\textbf {\bibinfo
  {volume} {576}},\ \bibinfo {pages} {899} (\bibinfo {year}
  {2002})}\BibitemShut {NoStop}%
\bibitem [{\citenamefont {{Baruteau}}\ \emph {et~al.}(2011)\citenamefont
  {{Baruteau}}, \citenamefont {{Cuadra}},\ and\ \citenamefont
  {{Lin}}}]{2011ApJ...726...28B}%
  \BibitemOpen
  \bibfield  {author} {\bibinfo {author} {\bibfnamefont {C.}~\bibnamefont
  {{Baruteau}}}, \bibinfo {author} {\bibfnamefont {J.}~\bibnamefont
  {{Cuadra}}},\ and\ \bibinfo {author} {\bibfnamefont {D.~N.~C.}\ \bibnamefont
  {{Lin}}},\ }\href {https://doi.org/10.1088/0004-637X/726/1/28} {\bibfield
  {journal} {\bibinfo  {journal} {\apj}\ }\textbf {\bibinfo {volume} {726}},\
  \bibinfo {eid} {28} (\bibinfo {year} {2011})}\BibitemShut {NoStop}%
\bibitem [{\citenamefont {{Yang}}\ \emph {et~al.}(2019)\citenamefont {{Yang}},
  \citenamefont {{Bartos}}, \citenamefont {{Haiman}}, \citenamefont {{Kocsis}},
  \citenamefont {{M{\'a}rka}}, \citenamefont {{Stone}},\ and\ \citenamefont
  {{M{\'a}rka}}}]{2019ApJ...876..122Y}%
  \BibitemOpen
  \bibfield  {author} {\bibinfo {author} {\bibfnamefont {Y.}~\bibnamefont
  {{Yang}}}, \bibinfo {author} {\bibfnamefont {I.}~\bibnamefont {{Bartos}}},
  \bibinfo {author} {\bibfnamefont {Z.}~\bibnamefont {{Haiman}}}, \bibinfo
  {author} {\bibfnamefont {B.}~\bibnamefont {{Kocsis}}}, \bibinfo {author}
  {\bibfnamefont {Z.}~\bibnamefont {{M{\'a}rka}}}, \bibinfo {author}
  {\bibfnamefont {N.~C.}\ \bibnamefont {{Stone}}},\ and\ \bibinfo {author}
  {\bibfnamefont {S.}~\bibnamefont {{M{\'a}rka}}},\ }\href
  {https://doi.org/10.3847/1538-4357/ab16e3} {\bibfield  {journal} {\bibinfo
  {journal} {\apj}\ }\textbf {\bibinfo {volume} {876}},\ \bibinfo {eid} {122}
  (\bibinfo {year} {2019})}\BibitemShut {NoStop}%
\bibitem [{\citenamefont {{Bartos}}\ \emph
  {et~al.}(2017{\natexlab{b}})\citenamefont {{Bartos}}, \citenamefont
  {{Haiman}}, \citenamefont {{Marka}}, \citenamefont {{Metzger}}, \citenamefont
  {{Stone}},\ and\ \citenamefont {{Marka}}}]{2017NatCo...8..831B}%
  \BibitemOpen
  \bibfield  {author} {\bibinfo {author} {\bibfnamefont {I.}~\bibnamefont
  {{Bartos}}}, \bibinfo {author} {\bibfnamefont {Z.}~\bibnamefont {{Haiman}}},
  \bibinfo {author} {\bibfnamefont {Z.}~\bibnamefont {{Marka}}}, \bibinfo
  {author} {\bibfnamefont {B.~D.}\ \bibnamefont {{Metzger}}}, \bibinfo {author}
  {\bibfnamefont {N.~C.}\ \bibnamefont {{Stone}}},\ and\ \bibinfo {author}
  {\bibfnamefont {S.}~\bibnamefont {{Marka}}},\ }\href
  {https://doi.org/10.1038/s41467-017-00851-7} {\bibfield  {journal} {\bibinfo
  {journal} {Nature Commun.}\ }\textbf {\bibinfo {volume} {8}},\ \bibinfo {eid}
  {831} (\bibinfo {year} {2017}{\natexlab{b}})}\BibitemShut {NoStop}%
\bibitem [{\citenamefont {{Corley}}\ \emph {et~al.}(2019)\citenamefont
  {{Corley}}, \citenamefont {{Bartos}}, \citenamefont {{Singer}}, \citenamefont
  {{Williamson}}, \citenamefont {{Haiman}}, \citenamefont {{Kocsis}},
  \citenamefont {{Nissanke}}, \citenamefont {{M{\'a}rka}},\ and\ \citenamefont
  {{M{\'a}rka}}}]{2019arXiv190202797C}%
  \BibitemOpen
  \bibfield  {author} {\bibinfo {author} {\bibfnamefont {K.~R.}\ \bibnamefont
  {{Corley}}}, \bibinfo {author} {\bibfnamefont {I.}~\bibnamefont {{Bartos}}},
  \bibinfo {author} {\bibfnamefont {L.~P.}\ \bibnamefont {{Singer}}}, \bibinfo
  {author} {\bibfnamefont {A.~R.}\ \bibnamefont {{Williamson}}}, \bibinfo
  {author} {\bibfnamefont {Z.}~\bibnamefont {{Haiman}}}, \bibinfo {author}
  {\bibfnamefont {B.}~\bibnamefont {{Kocsis}}}, \bibinfo {author}
  {\bibfnamefont {S.}~\bibnamefont {{Nissanke}}}, \bibinfo {author}
  {\bibfnamefont {Z.}~\bibnamefont {{M{\'a}rka}}},\ and\ \bibinfo {author}
  {\bibfnamefont {S.}~\bibnamefont {{M{\'a}rka}}},\ }\href@noop {} {\bibfield
  {journal} {\bibinfo  {journal} {arXiv:1902.02797}\ } (\bibinfo {year}
  {2019})}\BibitemShut {NoStop}%
\bibitem [{\citenamefont {{McKernan}}\ and\ \citenamefont
  {{Ford}}(2015)}]{2015MNRAS.452L...1M}%
  \BibitemOpen
  \bibfield  {author} {\bibinfo {author} {\bibfnamefont {B.}~\bibnamefont
  {{McKernan}}}\ and\ \bibinfo {author} {\bibfnamefont {K.~E.~S.}\ \bibnamefont
  {{Ford}}},\ }\href {https://doi.org/10.1093/mnrasl/slv076} {\bibfield
  {journal} {\bibinfo  {journal} {\mnras}\ }\textbf {\bibinfo {volume} {452}},\
  \bibinfo {pages} {L1} (\bibinfo {year} {2015})}\BibitemShut {NoStop}%
\bibitem [{\citenamefont {{Meiron}}\ \emph {et~al.}(2017)\citenamefont
  {{Meiron}}, \citenamefont {{Kocsis}},\ and\ \citenamefont
  {{Loeb}}}]{2017ApJ...834..200M}%
  \BibitemOpen
  \bibfield  {author} {\bibinfo {author} {\bibfnamefont {Y.}~\bibnamefont
  {{Meiron}}}, \bibinfo {author} {\bibfnamefont {B.}~\bibnamefont {{Kocsis}}},\
  and\ \bibinfo {author} {\bibfnamefont {A.}~\bibnamefont {{Loeb}}},\ }\href
  {https://doi.org/10.3847/1538-4357/834/2/200} {\bibfield  {journal} {\bibinfo
   {journal} {\apj}\ }\textbf {\bibinfo {volume} {834}},\ \bibinfo {eid} {200}
  (\bibinfo {year} {2017})}\BibitemShut {NoStop}%
\bibitem [{\citenamefont {{Inayoshi}}\ \emph {et~al.}(2017)\citenamefont
  {{Inayoshi}}, \citenamefont {{Tamanini}}, \citenamefont {{Caprini}},\ and\
  \citenamefont {{Haiman}}}]{2017PhRvD..96f3014I}%
  \BibitemOpen
  \bibfield  {author} {\bibinfo {author} {\bibfnamefont {K.}~\bibnamefont
  {{Inayoshi}}}, \bibinfo {author} {\bibfnamefont {N.}~\bibnamefont
  {{Tamanini}}}, \bibinfo {author} {\bibfnamefont {C.}~\bibnamefont
  {{Caprini}}},\ and\ \bibinfo {author} {\bibfnamefont {Z.}~\bibnamefont
  {{Haiman}}},\ }\href {https://doi.org/10.1103/PhysRevD.96.063014} {\bibfield
  {journal} {\bibinfo  {journal} {\prd}\ }\textbf {\bibinfo {volume} {96}},\
  \bibinfo {eid} {063014} (\bibinfo {year} {2017})}\BibitemShut {NoStop}%
\bibitem [{\citenamefont {{Wang}}\ \emph {et~al.}(2018)\citenamefont {{Wang}},
  \citenamefont {{Leigh}}, \citenamefont {{Yuan}},\ and\ \citenamefont
  {{Perna}}}]{2018MNRAS.475.4595W}%
  \BibitemOpen
  \bibfield  {author} {\bibinfo {author} {\bibfnamefont {Y.-H.}\ \bibnamefont
  {{Wang}}}, \bibinfo {author} {\bibfnamefont {N.}~\bibnamefont {{Leigh}}},
  \bibinfo {author} {\bibfnamefont {Y.-F.}\ \bibnamefont {{Yuan}}},\ and\
  \bibinfo {author} {\bibfnamefont {R.}~\bibnamefont {{Perna}}},\ }\href
  {https://doi.org/10.1093/mnras/sty107} {\bibfield  {journal} {\bibinfo
  {journal} {\mnras}\ }\textbf {\bibinfo {volume} {475}},\ \bibinfo {pages}
  {4595} (\bibinfo {year} {2018})}\BibitemShut {NoStop}%
\bibitem [{\citenamefont {{Gerosa}}\ and\ \citenamefont
  {{Berti}}(2017)}]{2017PhRvD..95l4046G}%
  \BibitemOpen
  \bibfield  {author} {\bibinfo {author} {\bibfnamefont {D.}~\bibnamefont
  {{Gerosa}}}\ and\ \bibinfo {author} {\bibfnamefont {E.}~\bibnamefont
  {{Berti}}},\ }\href {https://doi.org/10.1103/PhysRevD.95.124046} {\bibfield
  {journal} {\bibinfo  {journal} {\prd}\ }\textbf {\bibinfo {volume} {95}},\
  \bibinfo {eid} {124046} (\bibinfo {year} {2017})}\BibitemShut {NoStop}%
\bibitem [{\citenamefont {{Gerosa}}\ and\ \citenamefont
  {{Berti}}(2019)}]{2019arXiv190605295G}%
  \BibitemOpen
  \bibfield  {author} {\bibinfo {author} {\bibfnamefont {D.}~\bibnamefont
  {{Gerosa}}}\ and\ \bibinfo {author} {\bibfnamefont {E.}~\bibnamefont
  {{Berti}}},\ }\href@noop {} {\bibfield  {journal} {\bibinfo  {journal}
  {arXiv:1906.05295}\ } (\bibinfo {year} {2019})}\BibitemShut {NoStop}%
\bibitem [{\citenamefont {{Woosley}}(2017)}]{2017ApJ...836..244W}%
  \BibitemOpen
  \bibfield  {author} {\bibinfo {author} {\bibfnamefont {S.~E.}\ \bibnamefont
  {{Woosley}}},\ }\href {https://doi.org/10.3847/1538-4357/836/2/244}
  {\bibfield  {journal} {\bibinfo  {journal} {\apj}\ }\textbf {\bibinfo
  {volume} {836}},\ \bibinfo {eid} {244} (\bibinfo {year} {2017})}\BibitemShut
  {NoStop}%
\bibitem [{\citenamefont {{Belczynski}}\ \emph {et~al.}(2016)\citenamefont
  {{Belczynski}}, \citenamefont {{Heger}}, \citenamefont {{Gladysz}},
  \citenamefont {{Ruiter}}, \citenamefont {{Woosley}}, \citenamefont
  {{Wiktorowicz}}, \citenamefont {{Chen}}, \citenamefont {{Bulik}},
  \citenamefont {{O'Shaughnessy}}, \citenamefont {{Holz}}, \citenamefont
  {{Fryer}},\ and\ \citenamefont {{Berti}}}]{2016A&A...594A..97B}%
  \BibitemOpen
  \bibfield  {author} {\bibinfo {author} {\bibfnamefont {K.}~\bibnamefont
  {{Belczynski}}}, \bibinfo {author} {\bibfnamefont {A.}~\bibnamefont
  {{Heger}}}, \bibinfo {author} {\bibfnamefont {W.}~\bibnamefont {{Gladysz}}},
  \bibinfo {author} {\bibfnamefont {A.~J.}\ \bibnamefont {{Ruiter}}}, \bibinfo
  {author} {\bibfnamefont {S.}~\bibnamefont {{Woosley}}}, \bibinfo {author}
  {\bibfnamefont {G.}~\bibnamefont {{Wiktorowicz}}}, \bibinfo {author}
  {\bibfnamefont {H.~Y.}\ \bibnamefont {{Chen}}}, \bibinfo {author}
  {\bibfnamefont {T.}~\bibnamefont {{Bulik}}}, \bibinfo {author} {\bibfnamefont
  {R.}~\bibnamefont {{O'Shaughnessy}}}, \bibinfo {author} {\bibfnamefont
  {D.~E.}\ \bibnamefont {{Holz}}}, \bibinfo {author} {\bibfnamefont {C.~L.}\
  \bibnamefont {{Fryer}}},\ and\ \bibinfo {author} {\bibfnamefont
  {E.}~\bibnamefont {{Berti}}},\ }\href
  {https://doi.org/10.1051/0004-6361/201628980} {\bibfield  {journal} {\bibinfo
   {journal} {\aap}\ }\textbf {\bibinfo {volume} {594}},\ \bibinfo {eid} {A97}
  (\bibinfo {year} {2016})}\BibitemShut {NoStop}%
\bibitem [{\citenamefont {{Giacobbo}}\ and\ \citenamefont
  {{Mapelli}}(2018)}]{2018MNRAS.480.2011G}%
  \BibitemOpen
  \bibfield  {author} {\bibinfo {author} {\bibfnamefont {N.}~\bibnamefont
  {{Giacobbo}}}\ and\ \bibinfo {author} {\bibfnamefont {M.}~\bibnamefont
  {{Mapelli}}},\ }\href {https://doi.org/10.1093/mnras/sty1999} {\bibfield
  {journal} {\bibinfo  {journal} {\mnras}\ }\textbf {\bibinfo {volume} {480}},\
  \bibinfo {pages} {2011} (\bibinfo {year} {2018})}\BibitemShut {NoStop}%
\bibitem [{\citenamefont {{McKernan}}\ \emph {et~al.}(2019)\citenamefont
  {{McKernan}}, \citenamefont {{Ford}}, \citenamefont {{O'Shaughnessy}},\ and\
  \citenamefont {{Wysocki}}}]{2019arXiv190704356M}%
  \BibitemOpen
  \bibfield  {author} {\bibinfo {author} {\bibfnamefont {B.}~\bibnamefont
  {{McKernan}}}, \bibinfo {author} {\bibfnamefont {K.~E.~S.}\ \bibnamefont
  {{Ford}}}, \bibinfo {author} {\bibfnamefont {R.}~\bibnamefont
  {{O'Shaughnessy}}},\ and\ \bibinfo {author} {\bibfnamefont {D.}~\bibnamefont
  {{Wysocki}}},\ }\href@noop {} {\bibfield  {journal} {\bibinfo  {journal}
  {arXiv:1907.04356}\ } (\bibinfo {year} {2019})}\BibitemShut {NoStop}%
\bibitem [{\citenamefont {{Varma}}\ \emph {et~al.}(2019)\citenamefont
  {{Varma}}, \citenamefont {{Gerosa}}, \citenamefont {{Stein}}, \citenamefont
  {{H{\'e}bert}},\ and\ \citenamefont {{Zhang}}}]{2019PhRvL.122a1101V}%
  \BibitemOpen
  \bibfield  {author} {\bibinfo {author} {\bibfnamefont {V.}~\bibnamefont
  {{Varma}}}, \bibinfo {author} {\bibfnamefont {D.}~\bibnamefont {{Gerosa}}},
  \bibinfo {author} {\bibfnamefont {L.~C.}\ \bibnamefont {{Stein}}}, \bibinfo
  {author} {\bibfnamefont {F.}~\bibnamefont {{H{\'e}bert}}},\ and\ \bibinfo
  {author} {\bibfnamefont {H.}~\bibnamefont {{Zhang}}},\ }\href
  {https://doi.org/10.1103/PhysRevLett.122.011101} {\bibfield  {journal}
  {\bibinfo  {journal} {\prl}\ }\textbf {\bibinfo {volume} {122}},\ \bibinfo
  {eid} {011101} (\bibinfo {year} {2019})}\BibitemShut {NoStop}%
\bibitem [{\citenamefont {{Barausse}}\ \emph {et~al.}(2012)\citenamefont
  {{Barausse}}, \citenamefont {{Morozova}},\ and\ \citenamefont
  {{Rezzolla}}}]{2012ApJ...758...63B}%
  \BibitemOpen
  \bibfield  {author} {\bibinfo {author} {\bibfnamefont {E.}~\bibnamefont
  {{Barausse}}}, \bibinfo {author} {\bibfnamefont {V.}~\bibnamefont
  {{Morozova}}},\ and\ \bibinfo {author} {\bibfnamefont {L.}~\bibnamefont
  {{Rezzolla}}},\ }\href {https://doi.org/10.1088/0004-637X/758/1/63}
  {\bibfield  {journal} {\bibinfo  {journal} {\apj}\ }\textbf {\bibinfo
  {volume} {758}},\ \bibinfo {eid} {63} (\bibinfo {year} {2012})}\BibitemShut
  {NoStop}%
\bibitem [{\citenamefont {{Hofmann}}\ \emph {et~al.}(2016)\citenamefont
  {{Hofmann}}, \citenamefont {{Barausse}},\ and\ \citenamefont
  {{Rezzolla}}}]{2016ApJ...825L..19H}%
  \BibitemOpen
  \bibfield  {author} {\bibinfo {author} {\bibfnamefont {F.}~\bibnamefont
  {{Hofmann}}}, \bibinfo {author} {\bibfnamefont {E.}~\bibnamefont
  {{Barausse}}},\ and\ \bibinfo {author} {\bibfnamefont {L.}~\bibnamefont
  {{Rezzolla}}},\ }\href {https://doi.org/10.3847/2041-8205/825/2/L19}
  {\bibfield  {journal} {\bibinfo  {journal} {\apjl}\ }\textbf {\bibinfo
  {volume} {825}},\ \bibinfo {eid} {L19} (\bibinfo {year} {2016})}\BibitemShut
  {NoStop}%
\bibitem [{\citenamefont {{Farr}}\ \emph {et~al.}(2017)\citenamefont {{Farr}},
  \citenamefont {{Stevenson}}, \citenamefont {{Miller}}, \citenamefont {{Mand
  el}}, \citenamefont {{Farr}},\ and\ \citenamefont
  {{Vecchio}}}]{2017Natur.548..426F}%
  \BibitemOpen
  \bibfield  {author} {\bibinfo {author} {\bibfnamefont {W.~M.}\ \bibnamefont
  {{Farr}}}, \bibinfo {author} {\bibfnamefont {S.}~\bibnamefont {{Stevenson}}},
  \bibinfo {author} {\bibfnamefont {M.~C.}\ \bibnamefont {{Miller}}}, \bibinfo
  {author} {\bibfnamefont {I.}~\bibnamefont {{Mand el}}}, \bibinfo {author}
  {\bibfnamefont {B.}~\bibnamefont {{Farr}}},\ and\ \bibinfo {author}
  {\bibfnamefont {A.}~\bibnamefont {{Vecchio}}},\ }\href
  {https://doi.org/10.1038/nature23453} {\bibfield  {journal} {\bibinfo
  {journal} {\nat}\ }\textbf {\bibinfo {volume} {548}},\ \bibinfo {pages} {426}
  (\bibinfo {year} {2017})}\BibitemShut {NoStop}%
\bibitem [{\citenamefont {{Yi}}\ and\ \citenamefont
  {{Cheng}}(2019)}]{2019arXiv190908384Y}%
  \BibitemOpen
  \bibfield  {author} {\bibinfo {author} {\bibfnamefont {S.-X.}\ \bibnamefont
  {{Yi}}}\ and\ \bibinfo {author} {\bibfnamefont {K.~S.}\ \bibnamefont
  {{Cheng}}},\ }\href@noop {} {\bibfield  {journal} {\bibinfo  {journal}
  {arXiv:1909.08384}\ } (\bibinfo {year} {2019})}\BibitemShut {NoStop}%
\bibitem [{\citenamefont {{Bardeen}}(1970)}]{1970Natur.226...64B}%
  \BibitemOpen
  \bibfield  {author} {\bibinfo {author} {\bibfnamefont {J.~M.}\ \bibnamefont
  {{Bardeen}}},\ }\href {https://doi.org/10.1038/226064a0} {\bibfield
  {journal} {\bibinfo  {journal} {\nat}\ }\textbf {\bibinfo {volume} {226}},\
  \bibinfo {pages} {64} (\bibinfo {year} {1970})}\BibitemShut {NoStop}%
\bibitem [{\citenamefont {{Martini}}(2004)}]{2004cbhg.symp..169M}%
  \BibitemOpen
  \bibfield  {author} {\bibinfo {author} {\bibfnamefont {P.}~\bibnamefont
  {{Martini}}},\ }\href@noop {} {\bibfield  {journal} {\bibinfo  {journal}
  {Coevolution of Black Holes and Galaxies}\ ,\ \bibinfo {pages} {169}}
  (\bibinfo {year} {2004})},\ \Eprint {https://arxiv.org/abs/astro-ph/0304009}
  {astro-ph/0304009} \BibitemShut {NoStop}%
\bibitem [{\citenamefont {{Berti}}\ and\ \citenamefont
  {{Volonteri}}(2008)}]{2008ApJ...684..822B}%
  \BibitemOpen
  \bibfield  {author} {\bibinfo {author} {\bibfnamefont {E.}~\bibnamefont
  {{Berti}}}\ and\ \bibinfo {author} {\bibfnamefont {M.}~\bibnamefont
  {{Volonteri}}},\ }\href {https://doi.org/10.1086/590379} {\bibfield
  {journal} {\bibinfo  {journal} {\apj}\ }\textbf {\bibinfo {volume} {684}},\
  \bibinfo {pages} {822} (\bibinfo {year} {2008})}\BibitemShut {NoStop}%
\bibitem [{Note1()}]{Note1}%
  \BibitemOpen
  \bibinfo {note} {Hydrodynamical simulations by Secunda et al. (in prep)
  indicate aligned:anti-aligned ratios are likely in the 1:1 to 1:2
  range.}\BibitemShut {Stop}%
\bibitem [{\citenamefont {{Kimball}}\ \emph {et~al.}(2019)\citenamefont
  {{Kimball}}, \citenamefont {{Berry}},\ and\ \citenamefont
  {{Kalogera}}}]{2019arXiv190307813K}%
  \BibitemOpen
  \bibfield  {author} {\bibinfo {author} {\bibfnamefont {C.}~\bibnamefont
  {{Kimball}}}, \bibinfo {author} {\bibfnamefont {C.~P.}\ \bibnamefont
  {{Berry}}},\ and\ \bibinfo {author} {\bibfnamefont {V.}~\bibnamefont
  {{Kalogera}}},\ }\href@noop {} {\bibfield  {journal} {\bibinfo  {journal}
  {arXiv:1903.07813}\ } (\bibinfo {year} {2019})}\BibitemShut {NoStop}%
\bibitem [{\citenamefont {{Abbott}}\ \emph
  {et~al.}(2019{\natexlab{b}})\citenamefont {{Abbott}} \emph
  {et~al.}}]{2018arXiv181112940T}%
  \BibitemOpen
  \bibfield  {author} {\bibinfo {author} {\bibfnamefont {B.~P.}\ \bibnamefont
  {{Abbott}}} \emph {et~al.},\ }\href@noop {} {\bibfield  {journal} {\bibinfo
  {journal} {Astrophys. J. Lett.}\ }\textbf {\bibinfo {volume} {882}} (\bibinfo
  {year} {2019}{\natexlab{b}})}\BibitemShut {NoStop}%
\bibitem [{\citenamefont {{Chatziioannou}}\ \emph {et~al.}(2019)\citenamefont
  {{Chatziioannou}} \emph {et~al.}}]{2019arXiv190306742C}%
  \BibitemOpen
  \bibfield  {author} {\bibinfo {author} {\bibfnamefont {K.}~\bibnamefont
  {{Chatziioannou}}} \emph {et~al.},\ }\href@noop {} {\bibfield  {journal}
  {\bibinfo  {journal} {arXiv:1903.06742}\ } (\bibinfo {year}
  {2019})}\BibitemShut {NoStop}%
\bibitem [{\citenamefont {{Antonini}}\ \emph {et~al.}(2018)\citenamefont
  {{Antonini}}, \citenamefont {{Rodriguez}}, \citenamefont {{Petrovich}},\ and\
  \citenamefont {{Fischer}}}]{2018MNRAS.480L..58A}%
  \BibitemOpen
  \bibfield  {author} {\bibinfo {author} {\bibfnamefont {F.}~\bibnamefont
  {{Antonini}}}, \bibinfo {author} {\bibfnamefont {C.~L.}\ \bibnamefont
  {{Rodriguez}}}, \bibinfo {author} {\bibfnamefont {C.}~\bibnamefont
  {{Petrovich}}},\ and\ \bibinfo {author} {\bibfnamefont {C.~L.}\ \bibnamefont
  {{Fischer}}},\ }\href {https://doi.org/10.1093/mnrasl/sly126} {\bibfield
  {journal} {\bibinfo  {journal} {\mnras}\ }\textbf {\bibinfo {volume} {480}},\
  \bibinfo {pages} {L58} (\bibinfo {year} {2018})}\BibitemShut {NoStop}%
\bibitem [{\citenamefont {{Fishbach}}\ \emph {et~al.}(2017)\citenamefont
  {{Fishbach}}, \citenamefont {{Holz}},\ and\ \citenamefont
  {{Farr}}}]{2017ApJ...840L..24F}%
  \BibitemOpen
  \bibfield  {author} {\bibinfo {author} {\bibfnamefont {M.}~\bibnamefont
  {{Fishbach}}}, \bibinfo {author} {\bibfnamefont {D.~E.}\ \bibnamefont
  {{Holz}}},\ and\ \bibinfo {author} {\bibfnamefont {B.}~\bibnamefont
  {{Farr}}},\ }\href {https://doi.org/10.3847/2041-8213/aa7045} {\bibfield
  {journal} {\bibinfo  {journal} {\apjl}\ }\textbf {\bibinfo {volume} {840}},\
  \bibinfo {eid} {L24} (\bibinfo {year} {2017})}\BibitemShut {NoStop}%
\end{thebibliography}%

\end{document}